\documentclass[prd,nofootinbib,superscriptaddress]{revtex4}
\usepackage{amstext,amsmath,amssymb,amsfonts,bbm}
\usepackage{euscript}
\usepackage{color}

\def\be{\begin{equation}}
\def\ee{\end{equation}}
\def\bes{\begin{eqnarray}}
\def\ees{\end{eqnarray}}
\def\bay{\begin{array}}
\def\ear{\end{array}}

\def\pad{{\partial}}
\def\1{{{\mathbbm 1}}}
\def\bnab{{\mbox{\boldmath{$\nabla$}}}}
\def\anab{{\mbox{\sf\boldmath{{$\nabla$}}}}}
\def\half{\mbox{$1\over2$}}

\def\ra{{\rightarrow}}

\def\etal{\textit{et al.}}

\def\arb{\mathsf{b}}
\def\are{\mathsf{e}}
\def\af{{\sf f}}
\def\ak{{\sf k}}
\def\au{\mathsf{u}}

\def\sg{\mbox{\sl g}}

\def\bE{\mathbf{E}}

\def\bg{\mathbf{g}}
\def\boe{\mathbf{e}}
\def\bof{\mathbf{f}}
\def\bk{\mathbf{k}}

\def\hbb{{\hat{\mathbf{b}}}}
\def\hbe{{\hat{\boe}}}
\def\hbf{{\hat{\bof}}}
\def\hbk{{\hat{\bk}}}
\def\hbz{\hat{\mathbf{z}}}

\def\hmu{{\hat{\mu}}}
\def\hnu{{\hat{\nu}}}
\def\hrho{{\hat{\rho}}}

\def\eR{\EuScript{R}}

\def\co{{\cal O}}

\newcommand{\Omegab}{\mbox{\boldmath$\Omega$}}
\newcommand{\omegab}{\mbox{\boldmath$\omega$}}

\def\htheta{{\hat{\theta}}}
\def\hphi{{\hat{\phi}}}

\def\mout{{\mathrm{out}}}
\def\mrin{{\mathrm{in}}}

\def\ho{{\hat{1}} }
\def\htw{{\hat{2}} }
\def\hth{ {\hat{3}} }

\begin{document}
\title{Photon polarization and geometric phase in general relativity}
\author{Aharon Brodutch}
\affiliation{Department of Physics \& Astronomy, Macquarie University, Sydney NSW 2109, Australia}
\author{Tommaso F. Demarie}
\affiliation{Department of Physics \& Astronomy, Macquarie University, Sydney NSW 2109, Australia}
\author{Daniel R. Terno}
\affiliation{Department of Physics \& Astronomy, Macquarie University, Sydney NSW 2109, Australia}
\affiliation{Perimeter Institute for Theoretical Physics, 31 Caroline St. N.,  Waterloo ON N2L 2Y5, Canada}
\affiliation{Centre for Quantum Technologies, National University of Singapore, Singapore 117543}
\begin{abstract}
 Rotation of polarization in an external gravitational field is one of the effects of general relativity that
can serve as a basis for its precision tests. A careful analysis of reference frames is crucial for a proper
evaluation of this effect. We introduce an operationally-motivated local reference frame that allows for a
particularly simple description. We present a solution of null geodesics in Kerr space-time that is
organized around a new expansion parameter, allowing a better control of the series, and use it to
calculate the resulting polarization rotation. While this rotation depends on the reference-frame convention,
we demonstrate a gauge-independent geometric phase for closed paths in general space-times.

\end{abstract}
\maketitle

\section{Introduction}

Electromagnetic waves ---  visible light and other bands of the spectrum --- are our prime source of information about the Universe \cite{pad1}. Since measurements of the light deflection near the Sun \cite{e1919} were made  in 1919 wave propagation is used to test general relativity (GR). Two of the  ``classical tests" of GR, light deflection and time delay,  can be understood in terms of geometric optics \cite{will}.  The first post-eikonal approximation  \cite{bw} allows to track the evolution of electric and magnetic fields along the light rays and thus discuss polarization. 

In this approximation we  speak about photons with a null four-momentum $\ak$  and a transversal four-vector polarization $\af$. Both vectors are parallel-transported along the trajectory, which is a null geodesic \cite{mtw,chandra}:
\begin{align}
& \ak\!\cdot\ak=0, \qquad \anab_\ak\ak=0,  \label{momenone}\\
& \ak\!\cdot\af=0, \qquad \anab_\ak\af=0,  \label{polone}
 \end{align}
where $\anab_\ak$ is a covariant derivative along $\ak$.

In the last decades polarization has been yielding  important astrophysical and cosmological data. Cosmic microwave background \cite{cmb}, blazar flares \cite{fermi,fermilat}, astrophysical jets \cite{fermi,honda}, and  searches for dark matter \cite{dark-lens} are just several examples where polarization conveys crucial information.

Photons are  commonly used as physical carriers of quantum information. The abstract  unit of quantum information  is a qubit (a quantum bit) \cite{qinfo}, and   two linearly independent polarizations  encode the two basis states of a qubit.  One of the  branches of quantum information is quantum metrology, which aims to improve precision measurements by using explicitly quantum effects, such as entanglement \cite{macc}. Relativistic properties of the information carriers \cite{pt04,alsing09} become important when quantum technology is used in precision tests of relativity \cite{qtest} or quantum information processing  on the orbit \cite{milsat}.

Gravity causes polarization  to rotate. This effect is known as  a gravimagnetic/Faraday/Rytov-Skrotski\u{\i} rotation \cite{skrot,gravimag,fayos, kerr-farpol}. Helicity is invariant under rotations, but states of a definite helicity acquire  phases $e^{\pm i\Delta\chi}$. Depending on the context we  refer to $\Delta\chi$ either as a  polarization rotation or as a phase. Once evaluated this phase can be encapsulated as   a  quantum gate \cite{qinfo} and incorporated into (quantum) communication protocols or metrology tasks.

In the Schwarzschild space-time, as well as at the leading order of the post-Newtonian approximation, this phase is known to be zero \cite{skrot,pleb,god, gravimag}. It is  higher-order gravitational moments that are held responsible for the  rotation of the polarization plane. In particular, the GR effects were shown to dramatically alter  polarization of the X-ray radiation that is coming from the accretion disc of the (then presumed) black hole in Cyg X-1 \cite{bh:stark}. Numerous analytical and numerical studies of trajectories and polarizations in different models and astrophysical regimes were performed (\cite{gravimag,fayos,kerr-bray, kerr-farpol, lense, burns, kopeikin, faraoni08}, and references therein). The scenarios included fast-moving gravitating bodies, influence of gravitational lenses  and propagation through gravitational waves.

The results  are often contradictory. Some of the contradictions result from  genuine differences in superfluously similar physical situations \cite{fayos, burns, faraoni08}. On the other hand,  polarization rotation is operationally meaningful only if the evolving polarization vector is compared with some standard polarization basis (two linear polarizations, right- and left-circular polarizations, etc.) at each point along the ray \cite{lpt1, we-mach}.

Setting up  and aligning detectors requires alignment of local reference frames. Much work has been done recently on the role of reference frames in communications, especially in the context of quantum information \cite{Ref}. Partial knowledge of reference frames can lead to loss of communication capacity, and mistakes in identifying the information content of a physical system.    The lack of  definition for polarization standards and an \textit{ ad hoc} introduction of the angle adjustments is  one reason for the variety of  quoted values for the phase $\Delta\chi$. Even the consensual result of a zero phase in the  Schwarzschild space-time should be qualified. Without specifying the appropriate reference frame it is either meaningless or wrong.

In this article we investigate the role of local reference frames in defining $\Delta\chi$. The result is obviously gauge-dependent, and additional considerations should be used to fix the gauge.
Communicating reference frames is sometimes a difficult procedure which may involve a high communication cost. On a curved background of GR it may require some knowledge of the metric at each point along the trajectory. Using the method  presented by us in \cite{we-mach} we  build a local reference frame in stationary space-times, and then fix the standard polarizations in a single construction using what we call the {\it Newton gauge}.  Our construction does not require communication between the parties, gives a precise meaning to the idea that there is no polarization rotation in the Schwarzschild space-time and reproduces the absence of phase in the Minkowski space-time. From an operational point of view it allows us to set up detectors at any point in space based only on the local properties at that point.  We illustrate the use of this gauge  by studying polarization rotation in the Kerr space-time, and derive an explicit expression for $\Delta\chi$ in the scattering scenario that is described below.
The calculations are based on the results of \cite{kerr-bray,kerr-farpol,chandra} with the added value of the Newton gauge.  Careful bookkeeping is required with respect to the various coordinate systems used in different parts of the calculation, as well as different orders of the series expansion. To simplify the latter, we introduce a new  expansion parameter.

Closed paths result in a gauge-invariant gravity-induced phase. We discuss them in Section IV.
The rest of the paper is organized as follows: First we discuss null  trajectories in the Kerr space-time.
 In Section III we review the Wigner's construction of setting standard polarizations in a local frame, set-up the Newton gauge and explicitly calculate   the rotation in the scattering scenario, where the light is emitted and observed far from the gravitating body ($r_1,r_2\gg M,a$), but can pass close to it.  Finally, we discuss gauge-dependent and gauge-invariant aspects of our results  from mathematical and operational points of view (Section IV). Summary of the important facts about the Kerr space-time, as well as detailed calculations and special cases are presented in the appendices.

We use $-+++$ signature, set $G=c=1$ and use Einstein's summation convention in all dimensions. Three-dimensional  vectors are written in boldface and the unit vectors are distinguished  by carets, such as $\hbb$. Local tetrad components are written with carets on indices, such as $k^{\hat{\mu}}$, and the four-vector itself as $\ak=k^{\hat{\mu}}\are_{(\mu)}$, where $\are_{(\mu)}$ are vectors of  a local orthonormal tetrad.

\section{Null geodesics}
\subsection{Null geodesics in the Kerr space-time}
The Kerr metric in the Boyer-Lyndquist coordinates is given by \cite{ll2}
\be
ds^2=-\left(1-\frac{2M r}{\rho^2}\right)dt^2+\frac{\rho^2}{\Delta}dr^2+\rho^2d\theta^2+\left(r^2+a^2+\frac{2Mra^2}{\rho^2}\sin^2\!\theta\right)\sin^2\!\theta d\phi^2-\frac{4Ma r}{\rho^2}\sin^2\!\theta dt d\phi,
\ee
where $M$ is the mass and $a=J/M$ the angular momentum per unit mass of the gravitating body. We  use the standard notations
\be
\rho^2=r^2+a^2\cos^2\!\theta, \Delta=r^2-2Mr+a^2.
\ee

Thanks to the three conserved quantities --- the energy $E=-k_0$, the $z$-component of the angular momentum $L$ and the constant $\eta$ \cite{carter,chandra} --- the geodesic equations in Kerr space-time are integrable in quadratures. When dealing with photons it is convenient to set the energy to unity and to re-scale other quantities \cite{chandra}.  Explicit form and the asymptotic expansions of  $D=L/E$ and  $\eta$ are given in Appendix \ref{kerrdetail}.
Fixing the energy and using the null vector condition leaves us  with only two components of the four-momentum $\ak$. In specifying the initial data we usually take $k^\theta_1$ and $k^\phi_1$  as independent. 

 We discuss polarization from the point of view of static observers that are at rest in the ``absolute" space $t=\mathrm{const}$. At every point (outside the ergosphere, if the model represents a black hole) we introduce a chronometric orthonormal  frame \cite{fronov,mtw} (Appendix \ref{kerrdetail}), which will be used to express the initial conditions and observed quantities.

The Hamilton-Jacobi equation for a null geodesic separates and the trajectories can be deduced from it \cite{chandra}. We label their initial and final points as  $(r_1,\theta_1,\phi_1)$ and $(r_2,\theta_2,\phi_2)$, respectively. The gauge convention that we adapt  (Sec.~\ref{secnewton}) makes the knowledge of $\Delta\phi$ redundant for discussions of polarization rotation in the scattering scenario (For the calculation of $\phi$  see Appendix \ref{kerrdetail}).  In particular we are interested in the integrals
\be
\eR:=\int^r\!\frac{dr}{\pm\sqrt{R}}=\int^\theta\frac{d\theta}{\pm\sqrt{\Theta}}, \label{ert}
\ee
where
\begin{align}
&  R=r^4+(a^2-D^2-\eta)r^2+2M\bigl(\eta+(D-a)^2\bigr)r-a^2\eta, \\
&  \Theta=\eta+a^2\cos^2\!\theta-D^2\cot^2\theta.
\end{align}

  Null geodesics in Kerr space-time are classified according to the sign of the constant $\eta$. The scattering scenario corresponds to $\eta>0$, and typically $\eta\sim r_1^2$.
Following \cite{kerr-bray} we also assume that the angle $\theta$ reaches its extremal value (either maximum or minimum) only once, $\theta_1\rightarrow\theta_{\max\!/\min}\rightarrow\theta_2$. Series expansions of the above integrals are most conveniently written with the help of a new constant
\be
\Lambda^2:=D^2+\eta.
\ee
We use the parametrization $D=\Lambda \cos\alpha$, and  $\eta=\Lambda^2\sin^2\!\alpha$, with $0\leq\alpha\leq \pi$. In this notation the minimal coordinate distance from the center is
\be
r_{\min}=\Lambda -M -\frac{3M^2}{2\Lambda}+\frac{2a M \cos\alpha}{\Lambda}-\frac{a^2 \cos^2\!\alpha}{2 \Lambda}+\co(\Lambda^{-2}).
\ee

We perform the  integration
by the methods of \cite{kerr-bray,chandra} {(Appendix  B)}. In the scattering scenario $r_1, r_2\rightarrow\infty$, while the constants of motion are kept finite.
Expansion in terms of $\Lambda$  gives
\be
\eR=\frac{1}{\sqrt{\Lambda^2-a^2}}\left(\psi_1+\psi_2-\frac{a^2\sin^2\!\alpha}{4\Lambda^2}\big(3(\psi_1+\psi_2)+\sin\psi_1\cos\psi_1
+\sin\psi_2\cos\psi_2)\big)+\co(\Lambda^{-4})\right). \label{rte}
\ee
Where $\cos\psi=\cos\theta/\mu_+$ and $\mu_+$ is defined in the appendix.

 To separate  conceptual issues  from the computational details  we make two simplifying assumptions. First,  the initial and final points are taken to lie in the asymptotically flat regions, $r_i\rightarrow\infty$, while  $\eta$ and $D$ are kept finite. Second, in these regions the polar angle $\theta$ (nearly) reaches its asymptotic values $\theta_\mrin$ and $\theta_\mout$, respectively.

 The asymptotic form of momentum can be deduced from the equations of motion \cite{chandra}. In the  chronometric tetrad basis the outgoing momentum tends to
\be
k^{\hat{\mu}}\rightarrow\left(1,1,-\frac{s\beta_\mout}{r},\frac{D}{r \sin\theta_\mout}\right),
\ee
where $\beta_\mout:=+\sqrt{\Theta(\theta_\mout)}$ and $s=\pm 1$. This expression is true in any scenario where an outgoing photon reaches the asymptotically flat region.

The asymptotic expression for the incoming momentum is similar,
\be
k^{\hat{\mu}}\rightarrow\left(1,-1,\frac{s\beta_\mrin}{r},\frac{D}{r \sin\theta_\mrin}\right),
\ee
where  $\beta_\mrin:=+\sqrt{\Theta(\theta_\mrin)}$, and $s=\pm 1$ correspond to $\theta_\mrin\rightarrow\theta_{\max}/\theta_{\min}\rightarrow\theta_\mout$, respectively.

 Equating the radial and angular expressions for $\eR$ we obtain
\be
\theta_\mout=\pi-\theta_\mrin\pm\frac{4M}{\Lambda}\frac{\sqrt{\sin^2\!\alpha-\cos^2\!\theta_\mrin}}{\sin\theta_\mrin}+\co(\Lambda^{-2}),
\ee
with the higher order terms and special cases  described in Appendix \ref{kerrtraj}. The plus sign corresponds to the trajectory in which $\theta$ first decreases with $r$ and reaches $\theta_{\min}$ before increasing to $\theta_\mout$.

\subsection{Null geodesics in 1+3 formalism}
In stationary space-times  a tetrad of a static observer is naturally related to  Landau-Lifshitz 1+3 formalism \cite{ll2}. 
Static observers follow the congruence of time-like Killing vectors that  defines a projection from the  space-time manifold $\cal{M}$ onto a three-dimensional space $\Sigma_3$, $\pi:\mathcal{M}\rightarrow \Sigma_3$.

In practice it is performed by dropping the time-like coordinate of an event, and  vectors are projected by a push-forward map
$\pi_*\ak=\bk$ in the same way.
Contravariant vector components satisfy
\be
(\bk)^m\equiv(\pi_*\ak)^m=(\ak)^m=dx^m\!/d\lambda, \qquad m=1,2,3.
\ee
where $\lambda$ is the affine parameter.

The metric $\sg$ on $\mathcal{M}$ can be written in terms of a three-dimensional scalar $h$, a vector $\mathbf{g}$, and a metric $\gamma$ on $\Sigma_3$ as
\be
ds^2=-h(dx^0-\sg_mdx^m)^2+dl^2,
\ee
where the the three-dimensional distance is given by   $dl^2=\gamma_{mn}dx^mdx^n$. The metric components are
\be
\gamma_{mn}=\left(\sg_{mn}-\frac{\sg_{0m}\sg_{0n}}{\sg_{00}}\right),
\ee
and
\be
h=-\sg_{00}, \qquad \sg_{m}=-\sg_{0m}/\sg_{00}.
\ee
 The inner product of three-vectors will always refer to this metric, $\bk\!\cdot\!\bof=\gamma_{mn}k^mf^n$. Vector products and differential operators are defined as appropriate dual vectors \cite{ll2}. Finally,   the spatial projection of a null geodesic has a length $l$ that  is related to the affine parameter  $\lambda$ as
\be
\left(\frac{dl}{d\lambda}\right)^2=\frac{k_0^2}{h}=\bk^2=: k^2.
\ee
For a static observer the three spatial basis vectors of the local orthonormal tetrad are projected  into an orthonormal triad, $\pi_*\are_{(m)}= \hbe_{(m)}$, $\hbe_{(m)}\!\cdot\!\hbe_{(n)}=\delta_{mn}$.  We adapt a gauge  in which polarization  is orthogonal to the observer's four-velocity, $\au\!\cdot\!\af=0$. In a three-dimensional form this condition reads as  $\bk\!\cdot\!\bof=0$.

In stationary space-times the evolution equations Eq.~\eqref{momenone}, \eqref{polone} can be reduced into a convenient three-dimensional form \cite{fayos}.
Using the relations between four- and three-dimensional covariant derivatives, $\anab_\mu$ and $D_m$, respectively, the propagation equations \eqref{momenone} and  \eqref{polone} are brought to a three-dimensional form  \cite{fayos, gravimag},
\be
\frac{D\bk}{d\lambda}=\Omegab\times\bk +\frac{\bE_g\!\cdot\!\bk}{k^2}\bk, \qquad
\frac{D\hbf}{d\lambda}=\Omegab\times\hbf. \label{3deq}
\ee
The angular velocity of rotation $\Omegab$ is
\be
\Omegab=2\omegab-(\omegab\!\cdot\!\hbk)\hbk-\bE_g\times\bk,
\ee
where $\bg$ and $h$ play  the roles  of a vector potential of a gravimagnetic field $\mathbf{B}_g$,
\be
\omegab=-\half k_0\mathrm{curl}\ \!\mathbf{g}\equiv -\half k_0\mathbf{B}_g,
\ee
and a scalar potential of a gravielectric field,
\be
\bE_g=-\frac{\nabla h}{2 h},
\ee
respectively \cite{inertia}.

\section{Local reference frames and polarization rotation}

To discuss how photon's polarization rotates  it is necessary to define the standard polarization directions along its trajectory.  Only after physically defining two standard polarizations  it is meaningful to talk about polarization rotation, and only in a particular gauge the Schwarzschild space-time induces a zero phase on an open trajectory.  

\subsection{Wigner phase --- polarization convention}

Transversality of electromagnetic waves makes the choice of two standard  polarization directions  momentum-dependent.  On a curved background it  also depends on the location.
Wigner's construction of the massless representation of the Poincare group \cite{litgroup,wkt}  is the basis for classification of states in quantum field theory. We use it at every space-time point to produce  standard polarization vectors.  

The construction --- part of the induced representation of the Poincar\'{e} group \cite{wkt} --- consists of a choice of a standard reference momentum $\ak_S$ and two polarizations, a standard Lorentz transformation $L(\ak)$ that takes $\ak_S$ to an arbitrary momentum $\ak$, and a decomposition of an arbitrary Lorentz transformation $\Lambda$ in terms of the standard transformations and Wigner's little group element $W$. The standard reference 3-momentum is directed along  the $z$-axis of an arbitrarily chosen reference frame, with its $x$ and $y$ axes defining the  two linear polarization vectors $\hbb_{1,2}^S$, respectively. Hence $\ak_S=(1,0,0,1)$, and imposing the polarization gauge relates 3- and 4- polarization vectors, $\arb_{1,2}=(0,\hbb_{1,2})$.

The standard Lorentz transformation can be taken as \footnote{When it does not lead to confusion we use the same letter to label a four-dimensional object and its three-dimensional part. In particular,  $R$ stands both for a Lorentz transformation which is a pure rotation, and for the corresponding three-dimensional rotation matrix itself.}
\be
L(\ak)=R(\hbk)B_z(k),
\ee
 where $B_z(k)$ is a pure boost along the $z$-axis that takes  $\ak_S$ to $(k,0,0,k)$, and the standard rotation $R(\hbk)$ brings the $z$-axis to the desired direction $\hbk(\theta,\phi)$ first by rotating by $\theta$ around the $y$-axis and then by $\phi$ around the $z$-axis,
 \be
 R(\hbk)=R_z(\phi)R_y(\theta).
 \ee
The standard polarization vectors for an arbitrary momentum are defined as
\be
\hbb_i(\bk)= R(\hbk)\hbb_i^S,
\ee
and a general real 4-vector of polarization can be written as
\be
\af=\cos\chi\arb_1(\ak)+\sin\chi\arb_2(\ak). \label{poldec}
\ee

Helicity is invariant under Lorentz transformations.  The corresponding polarization vectors are
\be
\hbb_\pm(\bk)=\frac{1}{\sqrt{2}}\big(\hbb_1(\bk)\mp i \hbb_2(\bk)\big).
\ee
Under a Lorentz transformation $\Lambda$ a state of a definite helicity acquires a phase $e^{\pm i\xi(\Lambda,\bk)}$, which can be read-off from Wigner's little group element
\be
W(\Lambda,\ak)=L^{-1}(\Lambda\ak)\Lambda L(\ak)=R_z(\xi)T(\alpha,\beta).
\ee
The little group element $W$ leaves the standard momentum $\ak_S$ invariant. Here $T(\alpha, \beta)$ form a subgroup which is isomorphic to the translations of a Euclidean plane
and $R_z(\xi)$ is a rotation around the origin of that plane, which in this case is also a rotation
around the $z$-axis.

If the  transformation in question is a pure rotation $\eR$, then the little group element is then a  rotation $R_z(\xi)$, and the polarization three-vector is rotated by $\eR$ itself \cite{lpt1}. The phase $\chi$ has a simple geometric interpretation: it is the angle between, say, $\hbb_1(\eR\bk)$ and $\eR\hbb_1(\bk)$. It is zero if the standard polarizations of the new momentum are the same as the rotated standard polarizations of the old momentum.

This is what happens if a rotation $R_2(\omega)$  is performed around the current $\hbb_2(\bk)$: the resulting phase $\Delta\chi\big(R_2(\omega)\big)$ is zero \cite{we-mach}. Indeed, if $\hbk=\hbk(\theta,\phi)$, then by setting $\hbk'=R_2(\omega)\hbk$ and  using the decomposition
\be
R_2(\omega)=R(\hbk)R_y(\omega)R^{-1}(\hbk),
\ee
we find that $\hbk'=\hbk'(\theta+\omega,\phi)$, and the little group element is
\be
W=R^{-1}(\hbk')R_{2}(\omega) R(\hbk)=R_y^{-1}(\theta+\omega)R_y(\theta)R_y(\omega)=\1,
\ee
indicating the absence of rotation with respect to standard polarization basis.

In a curved space-time one has to provide the standard $(xyz)$ directions at every point. The next section deals with this problem.

\subsection{Newton gauge}\label{secnewton}

The three-dimensional propagation equations \eqref{3deq} in stationary space-times result in a joint rotation of polarization and unit tangent vectors
\be
\frac{D\hbk}{d\lambda}=\Omegab\times\hbk, \qquad
\frac{D\hbf}{d\lambda}=\Omegab\times\hbf. \label{3dprot}
\ee
In the Schwarzschild space-time test particles move in the plane  passing through the origin \cite{mtw,chandra,ll2},  $\Omegab=-\bE_g\times\bk$ and the covariant time-time component of the metric is related to the Newtonian gravitational potential $\varphi$ as $h=\sg_{00}=1+2\varphi(r)=1-2M/r$.
\textit{Requiring} the resulting phase to be zero, as  in \cite{skrot, gravimag,kopeikin}, constrains a choice of local reference frames.  We use the zero phase condition to  make a physically motivated choice of standard polarizations that does not require references to a parallel transport or communication between the observers.

Orient the local $z$-axis  along the direction of the free-fall acceleration, $\hbz\|\bnab\varphi$ and the standard polarizations as
\be
\hat{\mathbf{y}}\equiv\hbb_2:=\mathbf{w}\times\hbk/|\mathbf{w}\times\hbk|, \qquad \hbb_1:=\hbk\times\hbb_2, \label{polarbasis}
\ee
where $\bf{w}$ is a free-fall acceleration in the frame of a static observer. Then $\Omegab=\Omega\hbb_2$, and the propagation induces no  phase.

In a general static space-time we take the $z$-axis along the local free fall direction as seen by a static observer. In the Kerr space-time its components are \cite{fronov}
\be
w_{\hat{1}}=\frac{M(\rho^2-2r^2)\sqrt{\Delta}}{\rho^3(\rho^2-2Mr)}=-\frac{M}{r^2}-\frac{M^2}{r^3}+\co(r^{-4}),
\qquad w_{\hat{2}}=\frac{Mra^2\sin2\theta}{\rho^3(\rho^2-2Mr)}= a^2 M \frac{\sin2\theta}{r^4}+\co(r^{-5}), \qquad w_{\hat{3}}=0. \label{wcompo}
\ee

This convention, that we will call the Newton gauge, is consistent: if we set $\hbz=-\hat{\mathbf{r}}$ in the flat space-time, then  no phase is accrued as a result of the propagation. In addition to being defined by local operations, the Newton gauge has two further advantages. First, it does not rely on a weak field approximation to define the reference direction. Second,  if the trajectory is closed or self-intersecting, the reference direction $\hbz$ is the same at the points of the intersection.

\subsection{Kerr space-time examples}

 We impose the temporal gauge in local frames, hence
\be
f^{\hat{0}}\equiv 0 \Leftrightarrow f^t\equiv -\frac{2M a r f^\phi \sin^2\!\theta}{\rho^2- 2M r}=-\frac{2 Ma  f^{\hat{3}}\sin\theta }{r^2}+\co(r^{-3}), \label{mygauge}
\ee
where we used $f^{\hat{0}}=f^\mu e^{\hat{0}}_{(\mu)}$. This transversality reduces to a familiar three-dimensional expression $\ak\cdot\af\equiv\mathbf{k}\cdot\mathbf{f}= f^{\hat{1}}k^{\hat{1}}+f^{\hat{2}}k^{\hat{2}}+f^{\hat{3}}k^{\hat{3}}=0$. This also implies
\be
f_t= e_{(\mu)t} f^{\hat{\mu}}=e_{(0)t} f^{\hat{0}}=0.
\ee

In the Kerr space-time the Walker-Penrose quantity \cite{wp} $K_2+iK_1$,
\be
K_1=r B- a A \cos\theta, \qquad K_2= r A + a B \cos\theta, \label{wp12}
\ee
where
\begin{align}
A =&(k^t f^r-k^r f^t)+a(k^r f^\phi-k^\phi f^r)\sin^2\!\theta, \\
B=&\big( (r^2+a^2)(k^\phi k^\theta-k^\theta f^\phi)-a(k^t f^\theta-k^\theta f^t)\big)\sin\theta,
\end{align}
is conserved along null geodesics \cite{chandra}.  Transversality and the gauge \eqref{mygauge} make the Walker-Penrose constants functions of only two polarization  components, for example $f^{\hat{2}}$ and $f^{\hat{3}}$.

For a generic outgoing null geodesic in the asymptotic regime the constants become
\be
K_1=\gamma_\mout f^{\hat{2}}_\mout-s\beta_\mout f^{\hat{3}}_\mout, \qquad K_2=s\beta_\mout f^{\hat{2}}_\mout+\gamma_\mout f^{\hat{3}}_\mout,
\ee
where
\be
\gamma_\mout:=D\csc\theta_\mout-a\sin\theta_\mout,
\ee
and the analogous expression gives the constant in terms of the initial data. Hence,
\be
f^{\hat{1}}_\mout=0, \qquad f^{\hat{3}}_\mout=-\frac{1}{\beta^2_\mout+\gamma_\mout^2}(-s\beta_\mout K_1+\gamma_\mout K_2), \qquad f^{\hat{2}}_\mout=-\frac{1}{\beta^2_\mout+\gamma_\mout^2}(-s\beta_\mout K_2-\gamma_\mout K_1).
\ee

 To determine the polarization rotation $\chi$ we take the initial polarization to be, say, $\mathbf{f}_{\mrin}= \hbb_1^{\mrin}$, which often considerably simplifies the expression for $K_1$ and $K_2$.
Since $K_1$ and $K_2$ are linear functions of polarization, expressing the above result in the basis $(\hbb_1^{\mout},\hbb_2^{\mout})$ the desired  rotation angle $\chi$ is observed from
\be
\left(\begin{array}{c}
1 \\
0
\end{array}\right)
\rightarrow T\left(\begin{array}{c}
1 \\
0
\end{array}\right)=\left(\begin{array}{c}
\cos\chi \\
\sin\chi
\end{array}\right),
\ee
where $T$ is an orthogonal matrix that is described below.


We assume fixed $D$ and $\eta$. Using Eq.~\eqref{polarbasis} we find that at the limit $r_\mrin\ra\infty$ the initial standard linear polarization directions are
\be
\hbb_1^{\mrin}=\frac{1}{\sqrt{D^2+\beta_\mrin^2\sin^2\!\theta_\mrin}}\left(0,s\beta_\mrin\sin\theta_\mrin, D\right),
\qquad \hbb_2^{\mrin}=\frac{1}{\sqrt{D^2+\beta_\mrin^2\sin^2\!\theta_\mrin}}\left(0,D,-s\beta_\mrin\sin\theta_\mrin\right).
\ee
Similarly, the final standard polarizations are
\be
\hbb_1^{\mout}=\frac{1}{\sqrt{D^2+\beta_\mout^2\sin^2\!\theta_\mrin}}\left(0,s\beta_\mout\sin\theta_\mout, -D\right),
\qquad \hbb_2^{\mout}=\frac{1}{\sqrt{D^2+\beta_\mout^2\sin^2\!\theta_\mrin}}\left(0,D,s\beta_\mout\sin\theta_\mout\right).
\ee
The rotation matrices $N_{\mrin}$ and $N_\mout$ from the  polarization bases to $(\htheta,\hphi)$ basis are given in the Appendix C.

Following \cite{kerr-farpol} using the asymptotic relationship between Walker-Penrose constants and the components of polarization one can introduce a transformation matrix $R$,
\be
\left(\begin{array}{c}
f^{\hat{2}}_\mout \\
f^{\hat{3}}_\mout
\end{array} \right)=R\left(\begin{array}{c}
f^{\hat{2}}_\mrin \\
f^{\hat{3}}_\mrin
\end{array} \right).
\ee
The transformation matrix has the form
\be
R=\frac{1}{\sqrt{1+x^2}}\left(
\begin{array}{cc}
1 & -x \\
-x & -1
\end{array}
\right),
\ee
where the parameter $x$ is given by
\be
x=s\frac{\beta_\mrin\gamma_\mout-\beta_\mout\gamma_\mrin}{\gamma_\mrin\gamma_\mout+\beta_\mrin\beta_\mout},
\ee
with $\gamma_\mrin:=D\csc\theta_\mrin-a\sin\theta_\mrin$. Finally,
\be
T=N_\mout R N_\mrin^{-1}. \label{defT}
\ee
Expansion in the inverse powers of $\Lambda$ results in
\be
T_{21}=\sin\chi=-\frac{4Ma}{\Lambda^2}\cos\theta_\mrin +\co(\Lambda^{-3}). \label{mainf}
\ee
This is our main new result.  Few special cases are of interest. If the initial propagation  is parallel to the $z$-axis and the impact parameter equals $b$, then
\be
\Lambda^2=b^2-a^2, \qquad D=0,
\ee
and the polarization is rotated by
\be
\sin\chi=\frac{4Ma}{\Lambda^2}+\frac{15M^2a}{4\Lambda^3}+\co(\Lambda^{-4}),
\ee
and the antiparallel initial direction gives the opposite sign.

Motion in the equatorial plane corresponds to $\eta=0$ ($\sin\alpha=0$)  and is qualitatively similar to the motion in Schwarzschild space-time. If the trajectory starts there but eventually moves outside, then
 the polarization is rotated by
\be
\sin\chi=s\frac{8 M^2 a}{\Lambda^3}\sin\alpha +\co(\Lambda^{-4}).
\ee

In the case of initial propagation along the $z$-axis agrees with both \cite{skrot} and \cite{fayos}, if we take into account the respective definitions of reference frames. On the other hand, in a generic setting our Eq.~\eqref{mainf} differs by a power of $\Lambda$ from  \cite{kerr-farpol} or \cite{gravimag} (both works predict $\chi\sim\Lambda^{-3}$ but disagree on the pre-factors). The origin of this disagreement is in the following. In addition to using a more transparent series expansion, we use a different polarization basis from the one in \cite{kerr-farpol}. In particular, we do not require the knowledge of $\phi_\mout-\phi_\mrin$. Moreover,   \cite{gravimag} considered only a Machian effect (Sec.~\ref{phasesec}), which is  dominated by the reference-frame term \cite{we-mach}.

\section{Geometric phase} \label{phasesec}

A more geometric perspective on polarization rotation is possible both in a static space-time, where we use the 1+3 formalism, as well as in arbitrary space-times. While the phase $\Delta\chi$ is gauge-dependent for an open trajectory, we will show that it is gauge-invariant on a closed path.

Consider the differential equations for polarization rotation \cite{we-mach}.
  By setting $\af=\arb_1$ at the starting point
and  using the parallel transport equations Eqs.~\eqref{momenone} and \eqref{polone} we arrive to the equation
\be
\frac{d\chi}{d\lambda}=\frac{1}{\cos\chi}\anab_\ak\left(\af\!\cdot\!\arb_2\right)=
\frac{1}{\af\!\cdot\!\arb_1}\af\!\cdot\!\anab_\ak\arb_2.
\ee
 In a static space-time projection of the polarization on $\Sigma_3$ results in
\be
\frac{d\chi}{d\lambda}=\frac{1}{\cos\chi}\frac{D(\hbf\!\cdot\!\hbb_2)}{d\lambda},
\ee
hence the desired equation is
\be
\frac{d\chi}{d\lambda}=\frac{1}{\hbf\!\cdot\!\hbb_1}\left(\frac{D\hbf}{d\lambda}\!\cdot\! \hbb_2+\hbf\!\cdot\!\frac{D\hbb_2}{d\lambda}\right)=\omegab\!\cdot\!\hbk+\frac{1}{\hbf\!\cdot\!\hbb_2}\hbf
\!\cdot\!\frac{D\hbb_2}{d\lambda}. \label{defrot}
\ee
The first term corresponds to the original Machian effect that was postulated in \cite{god}, but the observable quantity involves both the Machian and the reference-frame terms \cite{we-mach}.

 Now we discuss the geometric meaning of these equations. First
 consider a  basis of 1-forms $(\sigma^1, \sigma^2, \sigma^3)$ that is dual to the orthonormal polarization basis $(\hbb_1, \hbb_2,\hbk)$ at every point of the trajectory.
  A matrix of connection 1-forms $\omega$ is written with the help of Ricci rotation coefficients $\omega^{\hat{\imath}}_{\hat{\jmath}\,\hat{l}}$ as $\omega^{\hat{\imath}}_{\ \hat{l}}=\omega^{\hat{\imath}}_{\hat{\jmath}\,\hat{l}}\sigma^j$ \cite{mtw, fra}, and a linear polarization is written as
\be
\hbf=f^{\hat{1}}\hbb_1+f^{\hat{2}}\hbb_2=\cos\chi\hbb_1+\sin\chi\hbb_2.
\ee
Its covariant derivative  equals to
\be
\frac{D\hbf}{d\lambda}=k\omega^{\hat{a}}_{\hat{3}\hat{c}}f^c\hbb_a+\omega^{\hat{3}}_{\hat{3}\hat{c}}f^c\bk+\bk(f^{\hat{c}})\hbb_c, \qquad a,c=1,2.
\ee
Taking into account the antisymmetry of the connections $\omega^{\hat{\imath}}_{\ \hat{l}}=\omega_{\hat{\imath} \hat{l}}=-\omega_{\hat{l} \hat{\imath}}$, we find
\be
\frac{D\bof}{d\lambda}=(-\hbb_1 f^{\hat{2}}+\hbb_2 f^{\hat{1}})\left(\frac{d\chi}{d\lambda}-\omega^{\hat{1}}_{\hat{3}\hat{2}}k\right)+
\hbk k(\omega^{\hat{3}}_{\hat{3}\hat{1}}f^{\hat{1}}+\omega^{\hat{3}}_{\hat{3}\hat{2}}f^{\hat{2}}).
\ee
A comparison with Eq. \eqref{3dprot} leads to the identification $\Omega^1=\Omegab\!\cdot\!\hbb_1=k\omega^{\hat{3}}_{\hat{3}\hat{2}}$, $\Omega^2=-k\omega^{\hat{3}}_{\hat{3}\hat{1}}$ and to an alternative equation for the polarization rotation,
\be
\frac{d\chi}{d\lambda}=\omegab\!\cdot\!\hbk+\omega^{\hat{1}}_{\hat{3}\hat{2}}k.
\ee
Given a trajectory with a tangent vector $\bk$  one can define a SO(2)  line bundle with the connection $\bar{\omega}=\omega^{\hat{1}}_{\hat{3}\hat{2}}kd\lambda$, similarly to the usual treatment of  geometric phase \cite{fra,geom}. Freedom of choosing the polarization frame $(\hbb_1,\hbb_2)$ at every point of the trajectory is represented by a SO(2) rotation $R_\hbk\big(\psi(\lambda)\big)$. Under its action the connection transforms as $\omega\rightarrow R\omega R^{-1}+R^{-1}dR$ \cite{fra}, so
\be
\frac{d\chi}{d\lambda}\rightarrow\omegab\!\cdot\!\hbk+\omega^{\hat{1}}_{\hat{3}\hat{2}}k+\frac{d\psi}{d\lambda}.
\ee

In a static space-time we can consider a closed trajectory in space. Then the resulting phase   is gauge-invariant, since the last term above is a total differential and drops out upon the integration on a closed contour,
 \be
 \Delta\chi=\oint \omegab\!\cdot\!\hbk d\lambda +\oint \bar{\omega}. \label{3dom}
 \ee
 We can formalize it with the help of a curvature 2-form is introduced as $\bar{\theta}:=d\bar{\omega}+\bar{\omega}\wedge \bar{\omega}.$ For the SO(2) bundle it reduces to $\bar{\theta}=d\bar{\omega}$, so by using Stokes' theorem the reference-frame term can be rewritten as a surface integral of the  bundle curvature  as
\be
\oint \bar{\omega}=\int\!\int \bar\theta.
\ee

A more practical expression follows from our previous discussion: $
\Delta\chi=\arcsin\hbf_\mout\!\cdot\!\hbb_2$.
Conservation of $K_1$ and $K_2$  in the Kerr space-time ensures that if a trajectory is closed as a result of the initial conditions, then $\hbf_\mout=\hbf_\mrin$ and $\Delta\chi=0$. The Newton gauge is designed to give a zero phase along any trajectory in the Schwarzschild space-time. As a result of the gauge invariance of Eq.~\eqref{3dom}, no gravitationally-induced phase is accrued along a closed trajectory in the Schwarzschild space-time, regardless of the gauge convention.

In a general space-time we introduce  an orthonormal tetrad such that $\ak=k\are_0+k\are_3$ at every point of the trajectory. We again impose a temporal gauge and set $\are_{1,2}=\arb_{1,2}$, where the local polarization basis is chosen according to some procedure. Then from Eq.~\eqref{poldec} it follows that
\begin{align}
\anab_\ak \af &= k^\hmu \mathsf{e}_{(\nu)} \omega_{\hmu \hrho}^\hnu f^\hrho + \frac{d f^\hnu}{d\lambda}\mathsf{e}_{(\nu)}
= \left( -f^{\hat{2}} \mathsf{e}_{(1)} + f^{\hat1} \mathsf{e}_{(2)}  \right) \left( \frac{d\chi}{d\lambda} -(\omega^{\hat1}_{\hat0 \hat2}+\omega^{\hat{1}}_{\hat3\hat2}) k \right)  \nonumber\\
&+\big(k(\omega^{\hat 0}_{\hat0\hat{c}}+\omega^{\hat0}_{\hat3\hat{c}})f^{\hat{c}}\big)\are_{(0)}+\big(k(\omega^{\hat 3}_{\hat0\hat{c}}+\omega^{\hat3}_{\hat3\hat{c}})f^{\hat{c}}\big)\are_{(3)}, \qquad c=1,2.
\end{align}
where we used again the antisymmetry of the connection. From the parallel transport condition $\anab_\ak \af = 0$, we can see that
\be
(\omega^{\hat 0}_{\hat0\hat{c}}+\omega^{\hat0}_{\hat3\hat{c}})f^{\hat{c}}=(\omega^{\hat 3}_{\hat0\hat{c}}+\omega^{\hat3}_{\hat3\hat{c}})f^{\hat{c}}=0.
\ee
and
\be
{d\chi} =(\omega^{\hat1}_{\hat0 \hat2}+\omega^{\hat{1}}_{\hat3\hat2}) k {d\lambda}:=\bar{\Omega}
\ee

We cannot have a closed trajectory in chronologically-protected space-time, so to obtain a gauge-invariant result we consider two future-directed trajectories that begin and end in the same space-time points. This layout is similar to the two arms of a Mach-Zender interferometer \cite{bw}. We also  align the initial and final propagation directions of the beams and use the same rules to define the standard polarizations. Similarly to the previous case the curvature  is  just $\bar{\Theta}=d\bar{\Omega}$, and the Stokes theorem gives the phase as
\be
\Delta\chi=\int_{\gamma_1}\bar{\Omega}-\int_{\gamma_2}\bar{\Omega}=\oint_\gamma \bar{\Omega}=\int\!\!\int\!\bar{\Theta}.
\ee

\section{Conclusions and outlook}
Rotation one ascribes to polarization   as light propagates on a curved background  depends on the gauge conventions that are used along the way.   However, the closed-loop phase is gauge-independent (assuming we use the same reference frame for transmission and detection).  Similarly to other instances of a geometric phase, the phase $\Delta\chi$ is given by the integral of the (bundle) curvature over the  surface that is bounded by the trajectory. The Newton gauge provides a convenient local definition of the standard polarizations. It is motivated by its relative simplicity and path-independence of the reference frames. It is also the gauge in which the statement of a zero accrued phase along an arbitrary path in the Schwarzschild space-time is correct.

 Our results characterize the  behavior of  polarization qubits on curved backgrounds. Specific regimes and realistic scenarios,  both on the Kerr background an beyond, are to be investigated farther. Of special interest are those scenarios that will be experimentally feasible in the near future, such as sending photons between satellites \cite{milsat}. Once the expression for $\Delta\chi$ is obtained, it can be represent it as the action  of a quantum gate \cite{qinfo}, similarly to the the special-relativistic scenarios \cite{pt04,bat05}. This will allow us to use the full toolbox of quantum optics to design  experiments. Understanding of the polarization phase in GR opens new possibilities for optics-based precision measurements \cite{we-mach}, both classical and quantum.

\acknowledgments

We thank P. Alsing, B.-L. Hu, A. Kempf, and N. Menicucci for discussions and comments.

 Research at Perimeter Institute is supported by the Government of Canada
through Industry Canada and by the Province of Ontario through the Ministry
of Research \& Innovation.

\appendix
\section{Some aspects of the Kerr space-time} \label{kerrdetail}

\subsection{Chronometric tetrad}
We fix an orthonormal tetrad at every space-time point (outside the static limit) by demanding that $\are_0$ is the four-velocity of a static observer, and the vectors $\are_1$ and $\are_2$ are proportional to the tangent vectors $\pad_r$ and $\pad_\theta$. Their covariant components $e_{(\alpha)\mu}$ are
\be
\are_{(0)}=\left(-\sqrt{1-\frac{2M r}{\rho^2}},0,0,-\frac{2M a r\sin^2\!\theta}{\rho\sqrt{\rho^2-2 Mr}}\right)
\ee
and
\be
\are_{(1)}=(0,\rho/\sqrt{\Delta},0,0), \qquad \are_{(2)}=(0,0,\rho,0), \qquad \are_{(3)}=\left(0,0,0,\rho\sin\theta\sqrt\frac{\Delta}{\Delta-a^2\sin^2\!\theta}\right)
\ee
Setting $E=-k_0=1$ leads to
\be
k^t=\frac{\rho^2-2M r a k^\phi \sin^2\!\theta}{\rho^2- 2M r},
\ee
and $k^r$ is expressed  from the null condition $\ak^2=0$,
 \be
(k^r)^2=\frac{\Delta[1-(k^\theta)^2(\Delta-a^2\sin^2\theta)-(k^\phi)^2\sin^2\theta\Delta]}{\Delta-a^2\sin^2\theta}.
\ee

 The spatial components of  the momentum in the coordinate and tetrad bases are related to $k^{\hat{m}}=e^{(m)}_{~~\mu} k^\mu$ as
\begin{align}
k^{\hat{1}}&=k^r\rho/\sqrt{\Delta},\\
k^{\hat{2}}&=k^\theta\rho,\\
k^{\hat{3}}&=k^\phi\rho\sin\theta\sqrt\frac{\Delta}{\Delta-a^2\sin^2\!\theta},
\end{align}
while $k^2=\bk^2=k^{\hat{m}}k_{\hat{m}}$ satisfies
\be
k^2=1+\frac{2Mr}{{\rho^2-2Mr}}=1+\frac{2M}{r}+\co(r^{-2}).
\ee
\subsection{Constants of motion}

The re-scaled ($E=1$) $z$-component of the angular momentum is
\be
D=\frac{1}{\rho^2-2Mr}\big(k^\phi(a^2+r^2)(\rho^2-2Mr)-2 Mar(1-k^\phi a \sin^2\!\theta)\big).
\ee
 The re-scaled  Carter's constant
\be
\eta:=K-(D-a)^2,
\ee
where the constant $K$ is most conveniently expressed as
\be
K:=(a \sin\theta-D/\sin\theta)^2+\big(\rho^2k^\theta\big)^2.
\ee
The asymptotic expressions when $r\rightarrow\infty$ and the momentum components are fixed are
\begin{align}
& D= k_{\hat{3}}r\sin\theta+\co(r^{-1}),\\
&\eta=\big(k_{\hat{3}}^2\cos^2\!\theta+k_{\hat{2}}^2\big)r^2 + a \cos^2\!\theta\big(a(k_{\hat{2}}^2+k_{\hat{3}}^2-1)-4Mk_{\hat{3}}\sin\theta\big)+\co(r^{-1}),\\
&\Lambda=r\sqrt{k_{\hat{2}}^2+k_{\hat{3}}^2}+\co(r^{-1}),\\
&\cos\alpha =\frac{k_{\hat{3}}\sin\theta}{\sqrt{k_{\hat{2}}^2+k_{\hat{3}}^2}}+\co(r^{-2}).
\end{align}

Scattering with the impact parameter $b$ that measures the coordinate distance $r\cos\theta$ from the $z$-axis is most conveniently  described with the help of a fiducial Cartesian system. Its  axes are ``parallel" to the fictitious Cartesian axes of the Byer-Lindquist coordinates, and in the asymptotic region it is just parallel to the global Cartesian grid. We introduce the momentum components $\hat{p}_i$ that are related to the spherical components $\hat{k}_i$ the the usual relations. In this case for the initial momentum parallel to the $z$-axis we have
\be
D=\co(r_1^{-3}), \qquad \Lambda=\sqrt{b^2-a^2}+\co(r_1^{-2}).
\ee

Some care is needed in treating  $R(r)$ as a function of the initial conditions when $r_1\rightarrow\infty$. Introducing the constants $c_0$, $c_1$, $c_2$, we write it as
\be\label{Rinr1}
R(r)=:r^4-c_2^2r_1^2 r^2+c_1 r_1^2 r+c_0 r_1^2,
\ee
where
\begin{align}
c_2^2 &:=(\Lambda^2-a^2)/r_1^2=(k_1^{\hat{2}})^2+(k_1^{\hat{3}})^2+\co(r_1^{-2}) \\
c_1 &:=2M(\Lambda^2+a^2-2\Lambda a\cos\alpha)/r_1=2 M (c_2^2-\frac{2ak_1^{\hat{2}}\sin\theta_1}{r_1})+\co(r_1^{-2}), \\
c_0 &:= -a^2\Lambda^2\sin^2\!\alpha=-a^2 \big((k_1^{\hat{2}})^2\cos^2\!\theta_1+(k_1^{\hat{3}})^2\big)+\co(r_1^{-2})
\end{align}

\section{Trajectories} \label{kerrtraj}


\subsection{The $r$ integral }\label{secrinteg}

 The  integral over $r$ in Eq. ~(\ref{ert}) is split into three parts,
\be
\eR:=\int^r\!\frac{dr}{\pm\sqrt{R}}=\eR_\infty-\eR_1-\eR_2,
\ee
where
\be
\eR_\infty:=2\int_{r_{\min}}^{\infty}\frac{dr}{\sqrt{R}}, \qquad \eR_i:=\int_{r_i}^\infty\frac{dr}{\sqrt{R}}. \label{rpart}
\ee

We present a corrected expression for $\eR_i$ and evaluate the term $\eR_\infty$ to the fourth order in $\Lambda$.
Starting from   $\eR_\infty$ in  using the method of \cite{kerr-bray} we
factor  out  $(r-r_{\min})$ and substite $r=r_{\min}/x$, writing $R(r)$ as
\be
R=\frac{r_{\min}^2}{x^4}\left(r_{\min}^2+Ax^2+\frac{Bx^3}{r_{\min}}+\frac{C x^4}{r_{\min}^2}\right)=:\frac{r_{\min}^2}{x^4}\tilde R(x),
\ee
where the constants $A$, $B$, $C$ are determined by the polynomial division.
Noting that $\tilde{ R}(1)=0$  we have
\be
 r_{\min}^2=-A-\frac{B}{r_{\min}}-\frac{C}{r_{\min}^2}
 \ee
 which gives
 \be
 \tilde R(x)=-A-\frac{B}{r_{\min}}-\frac{C}{r_{\min}}+Ax^2+\frac{B}{r_{\min}}+\frac{C x^4}{r_{\min}^2}
 \ee
Rewriting the equation and inserting the values for $A,B$ and $C$ gives us
\be
R(r)=\frac{r_{\min}^2}{x^4}(\Lambda^2-a^2)(1-x^2)\left(1+\frac{a^2\eta}{r_{\min}^2(\Lambda^2-a^2)}(1+x^2)
-\frac{2M}{r_{\min}}\frac{(D-a)^2+\eta}{\Lambda^2-a^2}\frac{1+x+x^2}{1+x}\right)
\ee
Defining
\be
f(x):=\frac{\eta(1+x^2)}{(\Lambda^2-a^2)},
\qquad g(x):=-2\frac{(D-a)^2+\eta}{\Lambda^2-a^2}\frac{1+x+x^2}{1+x},
\ee
we finally get
\be
\eR_\infty=2 \int^{\infty}_{r_{\min}}\!\!\frac{dr}{\sqrt{R}}=\frac{2}{\sqrt{\Lambda^2-a^2}}\int_0^1\!
dx\left((1-x^2)\left(1+\frac{a^2}{r_{\min}^2}f(x)+\frac{M}{r_{\min}}g(x)\right)\right)^{-\half}.
\ee
Expansion the integrand up to the third order in $\Lambda$ gives Eq.~(15).

\begin{align}
\eR_\infty&=\frac{1}{{\sqrt{\Lambda^2-a^2}}}\left(\pi+\frac{4M}{\Lambda}+
\frac{\pi}{\Lambda^2}\left(\frac{15}{4}M^2-\frac{3}{4}a^2\sin^2\!\alpha\right)-\frac{8 aM}{\Lambda^2}\cos\alpha \right.\nonumber \\
&+\left.\frac{M}{\Lambda^3}\left[\frac{128}{3}M^2+a^2(10 \cos^2\!\alpha-6\sin^2\!\alpha)+15\pi \label{rr} Ma\cos\alpha\right]\right)+\co(\Lambda^{-5}).
\end{align} \label{rinfo3}

The second and third integrals can be performed as an approximation in powers of $r_1$, using Eq.~(\ref{Rinr1}). Since
\be
R(r)=(r^4-c_2^2r_1^2r^2)\left (1+\frac{(c_0+c_1r)r_1^2}{r^2(r^2-c_2^2r_1^2)}\right ),
\ee
the leading terms in the expansion are
\be
\int_{r_i}^{\infty}\!\!\frac{dr}{\sqrt{R}}=\frac{1}{c_2r_1}\arcsin\frac{c_2r_1}{r_i}-\frac{c_1}{2}\left(\frac{2}{(c_2^2 r_1)^2}+\frac{2r_i^2-c_2^2r_1^2}{c_2^4 r_i r_1^2 \sqrt{r^2_i-c_2^2 r_1^2}}\right)+\co(r_1^{-5}),
\ee
with $r_i$ being either $r_1$ or $r_2$, where in the latter case we assume that $r_2\gtrsim r_1$.

\subsection{The $\theta$ integral}

We calculate the integral $\int1/\sqrt{\Theta}d\theta$ using the method of \cite{chandra}.  Noting that
\be
\Theta=\eta +(a-D)^2-(a\sin\theta-D\csc\theta)^2,
\ee
we perform a change of variables $\mu:=\cos\theta$ and obtain
\be
I_\theta=\int\!\!\frac{d\theta}{\sqrt{\Theta}}=\int\!\!\frac{d\mu}{\sqrt{\Theta_\mu}},
\ee
where fro $\eta>0$
\be
\Theta_\mu=a^2(\mu_-^2+\mu^2)(\mu_+^2-\mu^2), \qquad 0\leq\mu^2\leq\mu_+^2,
\ee
and
\be
\mu_\pm^2:=\frac{1}{2a^2}\left(\sqrt{(\Lambda^2-a^2)^2+4a^2\eta}\mp(\Lambda^2-a^2)\right).
\ee

The calculation is performed using the approximations to the auxiliary integral
\be\label{ellipticf}
\int_{\mu_*}^{\mu_+}\!\!\frac{d\mu}{\sqrt{\Theta_\mu}}=\frac{1}{a\sqrt{\mu_+^2+\mu_-^2}}F(\psi,k^2),
\ee
where $F(\psi,k^2)$ is  the elliptic integral of the first kind, and
\be
k^2:=\mu_+^2/(\mu_-^2+\mu_+^2), \qquad \cos\psi:=\mu_*/\mu_+,
\ee
with $0\leq\psi\leq\pi$ and $\mu_+=+\sqrt{\mu_+^2}$.

The asymptotic expansion in the powers of $\Lambda$ gives
\begin{align}
\mu_+^2 &=\sin^2\alpha+\frac{a^2}{4\Lambda^2}\sin^22\alpha+\co(\Lambda^{-4}) \\
\mu_-^2 &=\frac{\Lambda^2}{a^2}-\cos^2\alpha+\frac{a^2}{2\Lambda^2}\sin^22\alpha+\co(\Lambda^{-4})
\end{align}
so
\begin{align}
& k^2 = \frac{ a^2 \sin^2\alpha}{\Lambda^2}+ \co(\Lambda^{-4}),\\
& \frac{1}{a\sqrt{\mu_+^2+\mu_-^2}}=\frac{1}{\sqrt{\Lambda^2-a^2}}\left( 1-\frac{a^2}{\Lambda^2}\sin^2\alpha+\co(\Lambda^{-4})\right).
\end{align}

Trajectories of the type  $\theta_1\rightarrow\theta_{\min}/\theta_{\max}\rightarrow\theta_2$ correspond to $\cos\theta_{\min}=\mu_+$,  and $\cos\theta_{\max}=-\mu_+<0$, respectively. In both cases the integration leads to
\be
\eR=\frac{1}{a\sqrt{\mu_+^2+\mu_-^2}}\left[F(\psi_1,k^2)+F(\psi_2,k^2)\right],
\ee
and the expansion in powers of $\Lambda$ leads to Eq. (\ref{rte}).

 Using that at the zeroth order (the flat space-time) $\psi_1+\psi_2=\pi$, we expand $\psi_2=\pi-\psi_1+\delta_\psi$ 
as
\be
\psi_2=\pi-\psi_1 +\sum_k\frac{\xi_k}{\Lambda^k}.
\ee

Equating the two expression for $\eR_\infty$ in the scattering scenario leads to
\begin{align}
\xi_1&=4M, \\ \xi_2&=15 M^2 \pi/4-8 Ma\cos\alpha, \\
 \xi_3&= M(-6 a^2-128 M^2+45  Ma\pi\cos\alpha-24 a^2 \cos2\alpha
+a^2(3 +\cos2\psi_1) \sin^2\alpha).
\end{align}
Similarly, setting $\theta_2=\pi-\theta_1+\delta_\theta$,
\be
\delta_\theta=\sum_k\frac{\vartheta_k}{\Lambda^k},
\ee
  and expanding the both sides of
\be
\cos(\theta-\delta_\theta)=\mu_+\cos\left(\arccos\frac{\cos\theta_1}{\mu_+}-\delta_\psi\right),
\ee
we find
\be
\vartheta_1=4M\frac{\sqrt{\sin^2\!\alpha-\cos^2\!\theta_1}}{\sin\theta_1},
\ee

   In the special case of  scattering with the initial momentum parallel (anti-parallel) to the $z$-axis the angular momentum is zero, so $\mu_+=1$ and $\theta_\mrin=\psi_\mrin=0,\pi$, and it is easy to see that $\theta_\mout=\delta_\psi,\pi-\delta_\psi$, respectively.

\subsection{The $\phi$  motion}
	The  equation for $\phi$ is given by \cite{chandra}

\begin{align}
\phi= & \pm \left( D\int^r\!\frac{dr}{\sqrt{R}}+2Ma\int^r\!\frac{r dr}{\Delta \sqrt{R}}- a^2 D\int^r\!\frac{dr}{\Delta \sqrt{R}} \right) \nonumber \\
& \pm D\left(\int^\theta\!\frac{d\theta}{\sin^2\!\theta\sqrt{\Theta}}-\int^\theta\!\frac{d\theta}{\sqrt{\Theta}}\right).
\end{align}
If $\theta\neq\mathrm{const}$ the first and the last terms cancel thanks to Eq.~\eqref{ert}. For brevity we write the remaining terms as
\be
\phi_2-\phi_1=:
\pm\int^r\!\!R^\phi(r) dr \pm \int^\theta \!\!T(\theta)d\theta,
\ee

We again decompose the radial integral as $\eR^\phi=\eR^\phi_\infty-\eR^\phi_1-\eR^\phi_2$, where
\be
\eR^\phi_\infty=2\int_{r_{\min}}^\infty R^\phi(r) dr, \qquad \eR^\phi_i=\int^{\infty}_{r_i}\! R^\phi(r) dr.
\ee
Following the same procedure as in Sec.~B.1 we find
\be
\eR^\phi_\infty=\frac{a}{\sqrt{\Lambda^2-a^2}}\left(\frac{8M-a\pi\cos\alpha}{4\Lambda}+\frac{M(3M\pi-8 a\cos\alpha)}{6\Lambda^2}+\co(\Lambda^{-3})\right).
\ee
Expanding  $R^\phi(r)$ for $r\gg r_1$ we obtain
\begin{align}
R^\phi(r)&=\frac{2aM}{\sqrt{(r^2-c_2^2r_1^2)}\left (\sqrt{1+ \frac{c_1r_1^2}{r(r^2-c_2^2r_1^2)}+\frac{c_0r_1^2}{r^2(r^2-c_2^2r_1^2)}}\right)(r^2+a^2-2Mr)}+\\ \nonumber
&\frac{a^2D}{r\sqrt{(r^2-c_2^2r_1^2)}\left (\sqrt{1+ \frac{c_1r_1^2}{r(r^2-c_2^2r_1^2)}+\frac{c_0r_1^2}{r^2(r^2-c_2^2r_1^2)}}\right)(r^2+a^2-2Mr)} \nonumber \\
&=\frac{a}{r^2\sqrt{r^2-c_2^2r_1^2}}\left[2M\left(1+\frac{2M}{r}\right)+\frac{aD}{r}\right]+\co(r^{-5}),
\end{align}
where the coefficients $c_i$ are defined in Appendix A2. Hence
\be
\int_{r_i}^\infty\!\!R^\phi(r) dr=\frac{2aM}{r_i^2+r_i\sqrt{-c_2^2r_1^2+r_i^2}}
-\frac{a^2D}{2c_2^3r_1^3r_i^3}\left(c_2r_1\sqrt{-c_2^2r_1^2+r_i^2}-
r_i^2\arctan\left(\frac{c_2r_1}{\sqrt{-c_2^2r_1^2+r_i^2}}\right)\right)+\co(r_1^{-5}).
\ee

Integral $\int Td\theta$ leads to the elliptic integral of the third kind. A standard change of variables
$\cos\theta=\mu=\mu_+\cos t$ and the identity
\be
\sin^2\!\theta=(1-\mu_+^2)(1+p\sin^2\!t)
\ee
where
\be
p:=\frac{\mu_+^2}{1-\mu_+^2}=\tan^2\alpha\left(1+\frac{a^2}{\Lambda^2}\right)+\co(\Lambda^{-4})
\ee
lead to
\be
\int_{\theta_{\min}}^{\theta_*} \!\!T(\theta)d\theta=
\frac{1}{a\sqrt{\mu_+^2+\mu_-^2}}\frac{D}{1-\mu_+^2}\int_0^{\psi_*}\frac{dt}{(1+p\sin^2\!t)\sqrt{1-k^2\sin^2\!t}},
\ee
with $\cos\psi_*=\cos\theta_*/\mu_+$ as in Appendix B2.
Hence
\be
\int_{\theta_{\min}}^{\theta_*} \!\!T(\theta)d\theta=
\frac{\Lambda\cos\alpha}{a\sqrt{\mu_+^2+\mu_-^2}}\frac{1}{1-\mu_+^2}\Pi(-p,\psi_*,k^2),
\ee
where $\Pi(q,\psi,k^2)$ is the elliptic integral of the third kind.  Expanding the pre-factor in the powers of $\Lambda$ gives
\be
\frac{D}{a\sqrt{\mu_+^2+\mu_-^2}}\frac{1}{1-\mu_+^2}=\frac{\Lambda\sec\alpha}{\sqrt{\Lambda^2-a^2}}+\co(\Lambda^{-4}),
\ee
while
\be
\Pi(-p,\psi,0)=\frac{1}{\sqrt{1+p}}\left(\frac{\pi}{2}-\mathrm{arccot}\big(\sqrt{1+p}\tan\psi\big)\right)=:\Pi_0(p,\psi)
\ee
and
\be
\Pi(-p,\psi,k^2)=\Pi_0(p,\psi)+\frac{k^2}{2p}\left(\psi-\Pi_0(p,\psi)\right)+\co(k^4).
\ee
Taking into account that
\be
\frac{1}{\sqrt{1+p}}=|\cos\alpha|\left(1-\frac{a^2}{\Lambda^2}\sin^2\!\alpha\right)+\co(\Lambda^{-4}),
\ee
we get for the scattering scenario
\be
\phi_\mout-\phi_\mrin =\pi\frac{D}{|D|}+\sum_i\frac{\Phi_i}{\Lambda^i},
\ee
where the first two terms of the series are
\be
\Phi_1=\frac{\xi_1}{\cos^2\!\psi \cos\alpha + \sin^2\!\psi\sec\alpha},
\ee
and
\begin{align}
\Phi_2 &= \frac{ \xi_2 (3 + \cos\!2\alpha) \sec^3\!\alpha +
 4 \sec\alpha (\xi_1^2 \sin\!2\psi_1-\xi_2 \cos\!2\psi_1 ) \tan^2\alpha}{(1 + \sec^2\!\alpha - \cos 2\psi \tan^2\!\alpha)^2}
 +{\mbox{$1\over4$}}  a^2 \cos \alpha+ 2Ma,
\end{align}
where $\xi_i$ are given in Appendix B2.

\section{Newton gauge relationships} \label{gaugesec}
In this Appendix we present a relationship between the components of polarization in the chronometric tetrad and the Newton gauge basis. This relationship allows to calculate the Walker-Penrose conserved quantity  from the operationally meaningful polarization information and in the asymptotic regime gives the matrices $N$ of Eq.~\eqref{defT}.

The transversally of polarization $\hbk\cdot\hbf=0$ ensures the linearity of the relationship between $(f^{\hat2},f^{\hat3})$ and $(f^x,f^y)$, where
\be
\hbf=f^x\hbb_x+f^y\hbb_y.
\ee
Since ${\bf b}_y=\hat{\bf w}\times{\bf k}/|\hat{ \bf w}\times{\bf k}|$ and ${\bf b}_x=-{\bf b}_y\times\hat{\bf k}$, by using Eq.~\eqref{wcompo} we find
\begin{align}
&{\bf b}_y=\frac{1}{\cal{N}}\left(w_\htw{k}_\hth,-w_\htw{k}_\hth,-w_\htw{k}_\ho+w_\ho{k}_\htw\right),\\
&{\bf b}_x=\frac{1}{{\cal{N}}k}\left(w_\htw{k}_\ho {k}_\htw-w_\ho\big(k_\hth^2+{k}_\htw^2\big),\quad
w_\htw \big({k}_\htw^2+{ k}_\ho^2\big)-w_\ho { k}_\ho { k}_\htw,\quad -{k}_\hth(w_\ho{k}_\ho+w_\htw{k}_\hth)\right),\\
&{\cal N}^2=(w_\htw{k}_\hth)^2+(w_\htw{k}_\hth)^2+(-w_\htw{k}_\ho+a_\ho{k}_\htw)^2.
\end{align}
Projecting $\hbf $ on these directions gives
\begin{align}
&f^y=\frac{1}{{\cal{N}}{k}_\ho}\left(f^\htw(-w_\ho k_\ho k_\hth-w_\htw k_\hth  k_\htw)+f^\hth\big(w_\ho k_\ho k_\htw+a_\ho( k_\hth^2+ k_\ho^2)\big)\right),\\
&f^x=\frac{1}{{\cal{N}}{k}^\ho}\left(f^\htw(w_\htw k_\ho-w_\ho k_\htw)+f^\hth(-w_\ho k_\hth)\right).
\end{align}
We define the transformation matrix $N$ as
\begin{equation}
\left(\begin{array}{c}
f^\htw\\f^\hth
\end{array}\right)=N
\left(\begin{array}{c}
f^y\\f^x
\end{array}\right).
\end{equation}
In the asymptotic regime where $f^\ho\rightarrow0$ it becomes orthogonal.  Taking $k^{\hat{m}}=(1,s\beta/r,D/r \sin\theta)$ we obtain in the leading order $1/r$
\begin{align}
N=\frac{1}{\sqrt{\beta^2+D^2/\sin^2\!\theta}}\left(
\begin{array}{cc}
D/\sin\theta&s\beta\\
-s\beta&D/\sin\theta
\end{array}
\right).
\end{align}

\end{document}